\documentclass[aps,prd,reprint,groupedaddress,showpacs]{revtex4-1}
\usepackage{graphicx}
\usepackage{atlasphysics}
\usepackage{multirow}
\usepackage{subfigure}
\usepackage{amsmath}
\pdfoutput=1
\renewcommand{\pt}{$p_{\rm T}$}

\newcommand{\LamBDecay}{$\Lambda_b^0\to J/\psi(\mu^+\mu^-) \Lambda^0 (p\pi^-)$}
\newcommand{\LamBBarDecay}{$\bar\Lambda_b^0\to J/\psi(\mu^+\mu^-) \bar\Lambda^0 (\bar p\pi^+)$}
\newcommand{\BdDecay}{$B_d^0\to J/\psi(\mu^+\mu^-) K_S^0 (\pi^+\pi^-)$}

\newcommand{\Ks}{$K_S^0$}
\renewcommand{\Jpsi}{$J/\psi$}
\newcommand{\LamB}{$\Lambda_b^0$}
\newcommand{\LamBBar}{$\bar\Lambda_b^0$}

\newcommand{\Lam}{$\Lambda^0$}
\newcommand{\LamBar}{$\bar\Lambda^0$}

\newcommand{\Vz}{$V^0$}

\newcommand{\figWidth}{0.98}
\newcommand{\FirstImage}{top panel}
\newcommand{\SecondImage}{bottom panel}
\newcommand{\FIG}{Fig.}
\newcommand{\TABLE}{Table}
\newcommand{\BeginParams}{ }
\newcommand{\EndParams}{ }
\newcommand{\ItemParams}{ }
\newcommand{\BreakEqn}{\notag\\ & &  }
\newcommand{\paperTitle}{Measurement of the \boldmath{$\Lambda_b^0$} lifetime and mass in the ATLAS experiment}

\newcommand{\paperAbstract}{
  A measurement of the $\Lambda_b^0$ lifetime and mass in the decay channel $\Lambda_b^0\to J/\psi(\mu^+\mu^-) \Lambda^0 (p\pi^-)$ is presented. The analysis uses a signal sample of about 2200 $\Lambda_b^0$ and $\bar\Lambda_b^0$ decays that are reconstructed in 4.9~fb$^{-1}$ of ATLAS $pp$ collision data collected in 2011 at the LHC center-of-mass energy of 7~TeV. A simultaneous mass and decay time maximum likelihood fit is used to extract the $\Lambda_b^0$ lifetime and mass. They are measured to be $\tau_{\Lambda_b} = 1.449 \pm 0.036({\rm stat}) \pm 0.017({\rm syst})$~ps and $m_{\Lambda_b} = 5619.7 \pm 0.7({\rm stat}) \pm 1.1({\rm syst})$~MeV.
}
\usepackage{preprintcover}  % load package
\PreprintCoverPaperTitle{\paperTitle}  % paper title
\PreprintIdNumber{CERN-PH-EP-2012-163}  % CERN preprint number
\PreprintCoverAbstract{
  \paperAbstract
}  % paper abstract
\PreprintJournalName{Phys. Rev. D}  % journal name
\begin{document}

%%%%%%%%%%%%%%%%%%%%%%%%%%%%%%%%%%%%
%           Title page             %
%%%%%%%%%%%%%%%%%%%%%%%%%%%%%%%%%%%%

% Title
\title{Measurement of the \boldmath{$\Lambda_b^0$} lifetime and mass in the ATLAS experiment}

\author{The ATLAS Collaboration}
\date{\today}

%% -------------------------------------------------
%% Abstract
%% -------------------------------------------------
\begin{abstract}
  \paperAbstract
\end{abstract}

%% -------------------------------------------------
%% Title page
%% -------------------------------------------------

\maketitle

%%%%%%%%%%%%%%%%%%%%%%%%%%%%%%%%%%%%
%            Content               %
%%%%%%%%%%%%%%%%%%%%%%%%%%%%%%%%%%%%

\section{Introduction}
\label{section:intro}

The \LamB\ baryon (and its charge conjugate \LamBBar) is the lightest baryon containing a $b$ ($\bar b$) quark. With a mass of about $5620$~MeV \cite{PDG, Aaij2012241} it is not produced at $B$ factories, where the collision center-of-mass energy is tuned to produce pairs of $B$ mesons. Currently, hadron colliders are the only facilities where the properties of $b$ baryons can be studied. This paper presents a measurement of the \LamB\ mass and lifetime in the ATLAS experiment \cite{ATLAS} using the decay channel \LamBDecay\ (the charge conjugate mode is implied throughout the paper unless explicitly stated otherwise). The \LamB\ lifetime, although measured by many experiments \cite{PDG,PhysRevLett.104.102002,Aaltonen:2010pj,Abazov:2012iy}, still suffers from a large experimental uncertainty.

The decay \BdDecay\ has the same topology as the studied \LamB\ decay. The \Bd\ mass and lifetime are measured with good precision \cite{PDG}, and therefore this decay provides a useful tool to validate the \LamB\ results, as both measurements are subject to similar systematic uncertainties. The lifetime ratio, $\tau_{\Lambda_b}/\tau_{B_{d}}$, can be predicted by heavy quark expansion calculations \cite{HQET} and perturbative QCD \cite{NLO} and is of great theoretical interest. The lifetime and mass are determined using a simultaneous unbinned maximum likelihood fit to the reconstructed mass and decay time of each selected candidate.

%%%%%%%%%%%%%%%%%%%%%%%%%%%%%%%%%%%%%%%%%%%%%%%%%%%%%%%%%%%%%%%%%
\section{Data samples and trigger selection}

\label{section:data}
The ATLAS experiment \cite{ATLAS} is a general-purpose detector at the Large Hadron Collider (LHC). It covers nearly the entire solid angle around the interaction point with layers of tracking detectors, calorimeters, and muon chambers. The coordinate system has the $z$ axis aligned with the beam direction. The transverse momentum \pt and pseudorapidity, \eta, of reconstructed particles are defined with respect to that direction. This analysis uses two ATLAS subsystems: the inner detector (ID) and the muon spectrometer (MS). Both are situated in a magnetic field and serve as tracking detectors. The ID consists of three types of detector: the silicon pixel detector (Pixel), the silicon microstrip detector (SCT), and the transition radiation tracker (TRT). The MS consists of monitored drift tube chambers (MDT) and cathode strip chambers (CSC) for precision muon measurements, resistive plate chambers (RPC) and thin gap chambers (TGC) employed by the muon-trigger system. Tracks are reconstructed in the ID, and the MS is used to identify muons. Only tracks with \pt\ above 400~MeV and pseudorapidity $|\eta|<2.5$ are used in this analysis.

The analysis is based on data collected in 2011 using single-muon, dimuon, and \Jpsi\ triggers. The ATLAS trigger system \cite{Aad:2012xs} has three levels: the hardware-based Level-1 trigger and the two-stage High Level Trigger (HLT). At Level-1 the muon trigger uses dedicated fast muon-trigger chambers to search for patterns of hits corresponding to muons passing different \pt\ thresholds. Regions of interest (RoI) around these Level-1 hit patterns then serve as seeds for the HLT muon reconstruction. Since the rate from the low-\pt\ muon triggers was too high for all accepted events to be saved, prescale factors were applied to a subset of the triggers to reduce the output rate. The muon transverse momentum thresholds for single and dimuon triggers range from 4 to 22~GeV. The \Jpsi\ dimuon triggers require that the muons originate from a common vertex and have opposite charge, and that the dimuon mass is in the range $2.5\ {\rm GeV} < m_{\mu\mu} < 4.3$~GeV. The majority of the sample was collected by the \Jpsi\ trigger with a \pt\ threshold of 4~GeV applied to each muon. This was the lowest-\pt\ unprescaled trigger in the 2011 data taking; however, other complementary triggers were used, too. The \pt\ spectrum of the selected muons peaks at 5~GeV; the lowest muon \pt\ is above 2.5~GeV.

A Monte Carlo (MC) sample of $5 \times 10^6$ antibaryon \LamBBar\ events is used to study systematic effects and to correct for the efficiency and acceptance of the detector. The sample is generated using the {\sc Pythia~6} MC generator \cite{Pythia}  with the 2011 ATLAS AUET2B L0$^{**}$ tune \cite{tune}, and the events are filtered so that each accepted event has a decay \LamBBarDecay\ with the muons having transverse momenta of at least 2.5~GeV. The MC sample is generated with a \LamB\ lifetime of $\tau_{\Lambda_b}^{\rm MC} = 1.391$~ps.

%%%%%%%%%%%%%%%%%%%%%%%%%%%%%%%%%%%%%%%%%%%%%%%%%%%%%%%%%%%%%%%%%
\section{Reconstruction and signal selection}
\label{sec:reco}

\subsection{\boldmath{\Jpsi} and \boldmath{\Lam} preselection}

The decay \LamBDecay\ has a cascade topology. The \Jpsi\ decays instantly at the same point as the \LamB\ (secondary vertex), while the \Lam\ lives long enough to form a displaced tertiary vertex. There are four final-state particles: two muons from the \Jpsi, and a proton and a pion from the \Lam\ decay.

The dimuon and dihadron (\Vz) pairs are preselected by requiring that their tracks be successfully fitted to a common vertex satisfying some basic quality requirements. The \Jpsi\ and \Vz\ pre-selection is very loose, so that potential candidates are not excluded at this stage. The dimuon candidates are accepted if the \Jpsi\ vertex-refitted invariant mass lies in the range $2.8\ {\rm GeV} < m_{\mu\mu} < 3.4\ {\rm GeV}$. The dihadron candidates are accepted if the invariant mass is in the range $1.08\ {\rm GeV} < m_{p\pi} < 1.15\ {\rm GeV}$. The masses of a proton and a pion are assigned to the tracks when the invariant mass is calculated; $p\pi^-$ and $\bar p\pi^+$ combinations are tested so that both  \Lam\ and \LamBar\ candidates are accepted.

\subsection{Reconstruction of \boldmath{\LamBDecay}}
\label{section:lamb}

The muon and hadron track pairs preselected with the criteria described in the previous section are then refitted with a constraint of a \LamBDecay\ topology. The muons are constrained to intersect at a single vertex, while their invariant mass is set equal to the known mass of the \Jpsi, $m_{J/\psi} = 3096.92$~MeV \cite{PDG}. The two hadronic tracks are constrained to a second vertex, and their invariant mass is fixed to the mass of the \Lam, $m_{\Lambda^0} = 1115.68$~MeV \cite{PDG}. The combined momentum of the refitted \Vz\ track pair is constrained to point to the dimuon vertex in three dimensions. The fit is performed on all four tracks simultaneously, taking into account the constraints described above (cascade topology fit) and the full track error matrices. The quality of the fit is characterized by the value of $\chi^2/N_{\rm dof}$, where a global $\chi^2$ involving all four tracks is used. The corresponding number of degrees of freedom, $N_{\rm dof}$, is six. Furthermore, for each track quadruplet, that can be successfully fitted to the \LamB\ decay topology, a \BdDecay\ topology fit is attempted (i.e. a pion mass is assigned to the hadronic tracks and the \Vz\ mass is constrained to the mass of \Ks, $m_{K_S} = 497.65$~MeV \cite{PDG}). This is to identify possible $B_d^0$ decays mistaken for $\Lambda_b^0$.

The \LamB\ candidates are then subjected to the following selections:
\begin{itemize}
  \item The global $\chi^2/N_{\rm dof} < 3$.
  \item The transverse momentum of the cascade-refitted \Vz, $p_{{\rm T},V^0} > 3.5$~GeV.
  \item The transverse decay length of the cascade-refitted \Vz\ vertex measured from the \LamB\ vertex, $L_{xy,V^0} > 10$~mm.
  \item The invariant mass must be in the range $5.38\ {\rm GeV} < m_{J/\psi\Lambda^0} < 5.90\ {\rm GeV}$.
  \item If the four tracks forming a \LamB\ candidate also result in an acceptable \Bd\ fit, the candidate must have a difference of cumulative $\chi^2$ probabilities of the two fits, $\mathcal{P}_{\Lambda^0_b} - \mathcal{P}_{B_d} > 0.05$.
\end{itemize}

\begin{figure}[t]
\begin{center}
  \includegraphics[width=\figWidth\linewidth]{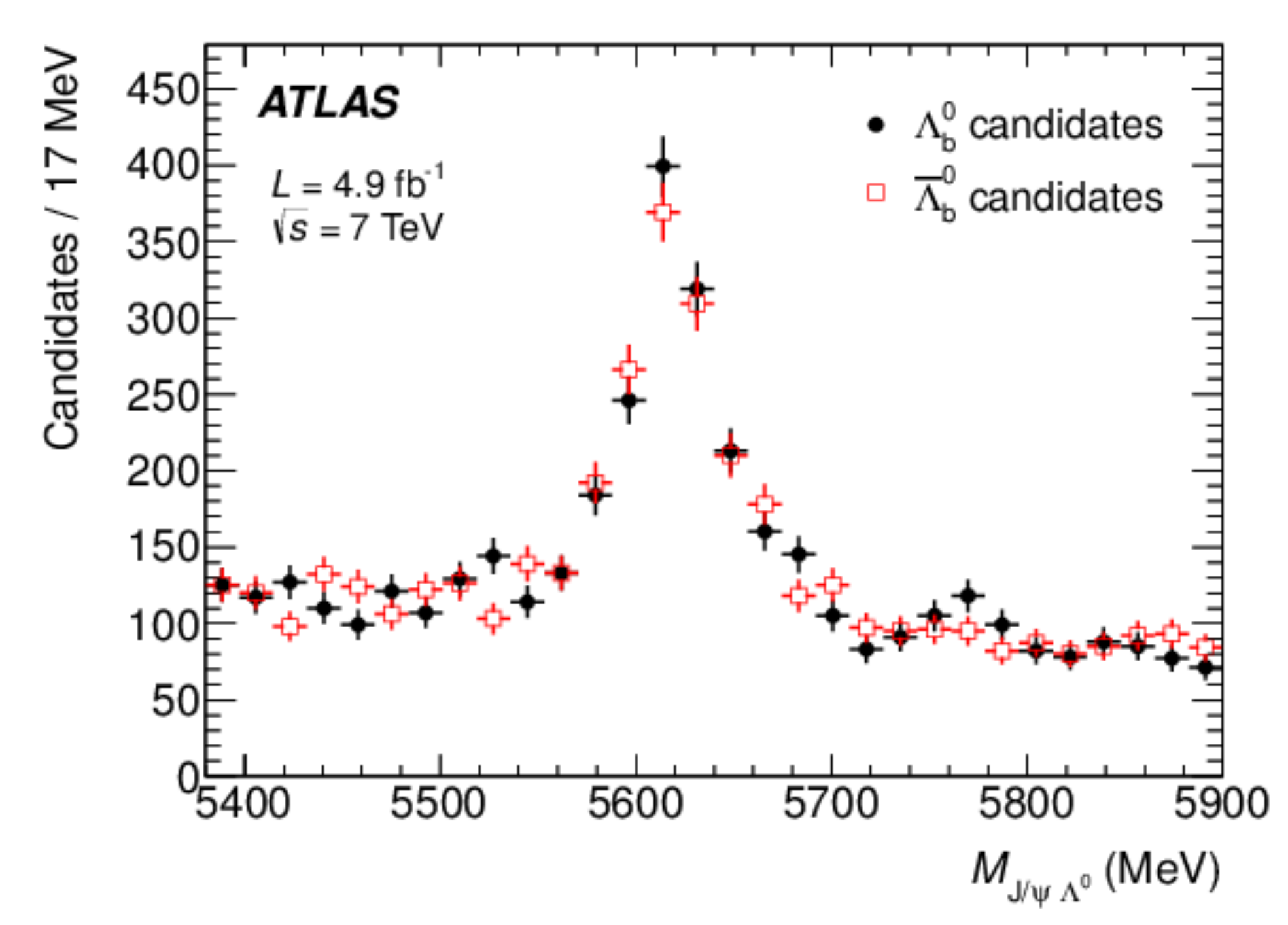}

  \caption{Invariant mass distribution of the selected \LamB\ and \LamBBar\ candidates.}

  \label{fig:LamB_LamBBar}
\end{center}
\end{figure}

With these criteria, 4074 \LamB\ and 4081 \LamBBar\ candidates (including background) are selected. No track quadruplet is successfully fitted as both a \LamB\ and a \LamBBar\ decay. The mass distributions of the selected candidates are shown in \FIG~\ref{fig:LamB_LamBBar}. In the rest of the paper the \LamB\ and \LamBBar\ samples are combined.

%%%%%%%%%%%%%%%%%%%%%%%%%%%%%%%%%%%%%%%%%%%%%%%%%%%%%%%%%%%%%%%%%
\section{Mass and proper decay time fit}
\label{sec:fit}

The proper decay time of the \LamB\ candidate is calculated from the measured decay distance and the candidate's momentum as follows:
\begin{equation}
  \tau = \frac{L_{xy} m^{\rm PDG}} {p_{\rm T}}, \label{eqn:tau} \notag
\end{equation}
where $m^{\rm PDG} = 5619.4$~MeV \cite{PDG}, $p_{\rm T}$ is the reconstructed \LamB\ transverse momentum, and $L_{xy}$ is the \LamB\ transverse decay distance measured from the primary vertex (PV). On average there are 6.8 collision vertices per event in the selected data resulting from multiple collisions at each LHC bunch crossing (pileup events). The collision vertex that in three-dimensional space lies closest  to the trajectory of the reconstructed \LamB\ candidate is used as the PV.

An unbinned maximum likelihood fit is used to determine the \LamB\ mass and lifetime. The mass and proper decay time are fitted using a likelihood function defined as follows:
\begin{eqnarray}
  L &=& \prod\limits_{i=1}^{N} \Big[ f_{\rm sig}      \mathcal{M}_{\rm s}(m_i | \delta_{m_i}) \mathcal{T}_{\rm s}(\tau_i | \delta_{\tau_i}) w_{\rm s}(\delta_{m_i},\delta_{\tau_i}) + \BreakEqn
                                   (1-f_{\rm sig})  \mathcal{M}_{\rm b}(m_i | \delta_{m_i}) \mathcal{T}_{\rm b}(\tau_i | \delta_{\tau_i}) w_{\rm b}(\delta_{m_i},\delta_{\tau_i}) \Big], \notag
\end{eqnarray}
where $f_{\rm sig}$ denotes the fraction of signal candidates; $m_i$ is the invariant mass of the $i$-th candidate and $\tau_i$ is its proper decay time. The corresponding errors, $\delta_{m_i}$ and $\delta_{\tau_i}$, are estimated on a candidate-by-candidate basis by the cascade topology fit. $\mathcal{M}_{\rm s}$ and $\mathcal{M}_{\rm b}$ are probability density functions (PDFs) describing the signal and background mass dependence; $\mathcal{T}_{\rm s}$ and $\mathcal{T}_{\rm b}$ describe the dependence on the proper decay time. The invariant mass and proper decay time error distributions, $w_{\rm s(b)}(\delta_{m},\delta_\tau) = w'_{\rm s(b)}(\delta_m) w''_{\rm s(b)}(\delta_\tau)$, are extracted from data. It has been verified that using separate PDFs for the signal and background component produces the same result when a single PDF is used, $w \equiv w_{\rm s} = w_{\rm b} $. For this reason the latter case is used.

The background can be divided into two categories: prompt and nonprompt backgrounds. The prompt background consists of \Jpsi\ candidates produced directly in the $pp$ collision that are randomly combined with \Vz\ candidates, which also include fake combinatorial \Lam\ or \Ks\ candidates. The prompt background decay length is due to the finite resolution of the vertex reconstruction. The nonprompt background includes events where the \Jpsi\ candidate originates in the decay of a $b$ hadron. This type of background has a lifetime due to its origin in long-lived $b$ hadrons [e.g. \BdDecay, with the \Ks\ meson misidentified as \Lam, forming a nonprompt background for \LamB].

The signal component of the mass PDF, $\mathcal{M}_{\rm s}$, is a Gaussian function with a mean equal to $m_{\Lambda_b}$ and width $S_m \delta_m$. The mass error scale factor $S_m$ determines how much the errors $\delta_{m_i}$ are overestimated or underestimated. The background component is a first order polynomial with a slope~$b$.

Using the estimated decay time error $\delta_\tau$, the proper decay time resolution is modeled with a Gaussian function:
\begin{equation}
  R(\tau - \tau'|\delta_\tau) = \frac{ 1 }{ \sqrt{2\pi} S_{\tau} \delta_\tau }\ e^{ -\frac{\left(\tau - \tau'\right)^2}{2\left(S_{\tau} \delta_\tau\right)^2} },  \label{eqn:res}
\end{equation}
where $S_{\tau}$ denotes the proper decay time error scale factor, and $\tau$ and $\tau'$ stand for the reconstructed and true proper decay times, respectively.

The signal and nonprompt background proper decay time distributions are modeled as exponential functions, $E(\tau'; \tau_B)$, for $\tau'>0$, with $\tau_B$ being the fitted parameter denoting either the \LamB\ lifetime or the pseudolifetime of the long-lived background. The prompt background component is modeled by a sum of two functions: a Dirac $\delta$ function $\delta_{\rm Dirac}(\tau')$ and a symmetric exponential (Laplace distribution) $E_{\rm sym}(\tau')$, to account for the non-Gaussian tails of the prompt background observed in data.

The functions are convolved with the resolution model (\ref{eqn:res}) to obtain the PDFs of the measured proper decay time:
\begin{eqnarray}
  \mathcal{T}_{\rm s}(\tau|\delta_\tau) &=&  \varepsilon(\tau')^{-1} E(\tau'; \tau_{\Lambda_b}) \otimes R(\tau - \tau'|\delta_\tau), \label{eqn:tauSignal}\\
  \mathcal{T}_{\rm b}(\tau|\delta_\tau) &=& \Big[ f_{1} \mathcal{T}_{\rm p}(\tau') +
                                                   (1-f_{1})\mathcal{T}_{\rm np}(\tau') \Big] \otimes  \BreakEqn
                                            R(\tau - \tau'|\delta_\tau),\notag
\end{eqnarray}
with the nonprompt and prompt components defined as
\begin{eqnarray}
  \mathcal{T}_{\rm np}(\tau') &=& f_2 E(\tau'; \tau_{\rm bkg,1}) + (1-f_2) E(\tau'; \tau_{\rm bkg,2}),\notag\\
  \mathcal{T}_{\rm p}(\tau')  &=& f_3     \delta_{\rm Dirac}(\tau') + (1-f_3) E_{\rm sym} (\tau'; \tau_{\rm bkg,3}).\notag
\end{eqnarray}
The efficiency correction function $\varepsilon(\tau')$ in Eq.~(\ref{eqn:tauSignal}) accounts for the decay-time-dependent selection bias.

\begin{figure}[t]
\begin{center}
  \includegraphics[width=\figWidth\linewidth]{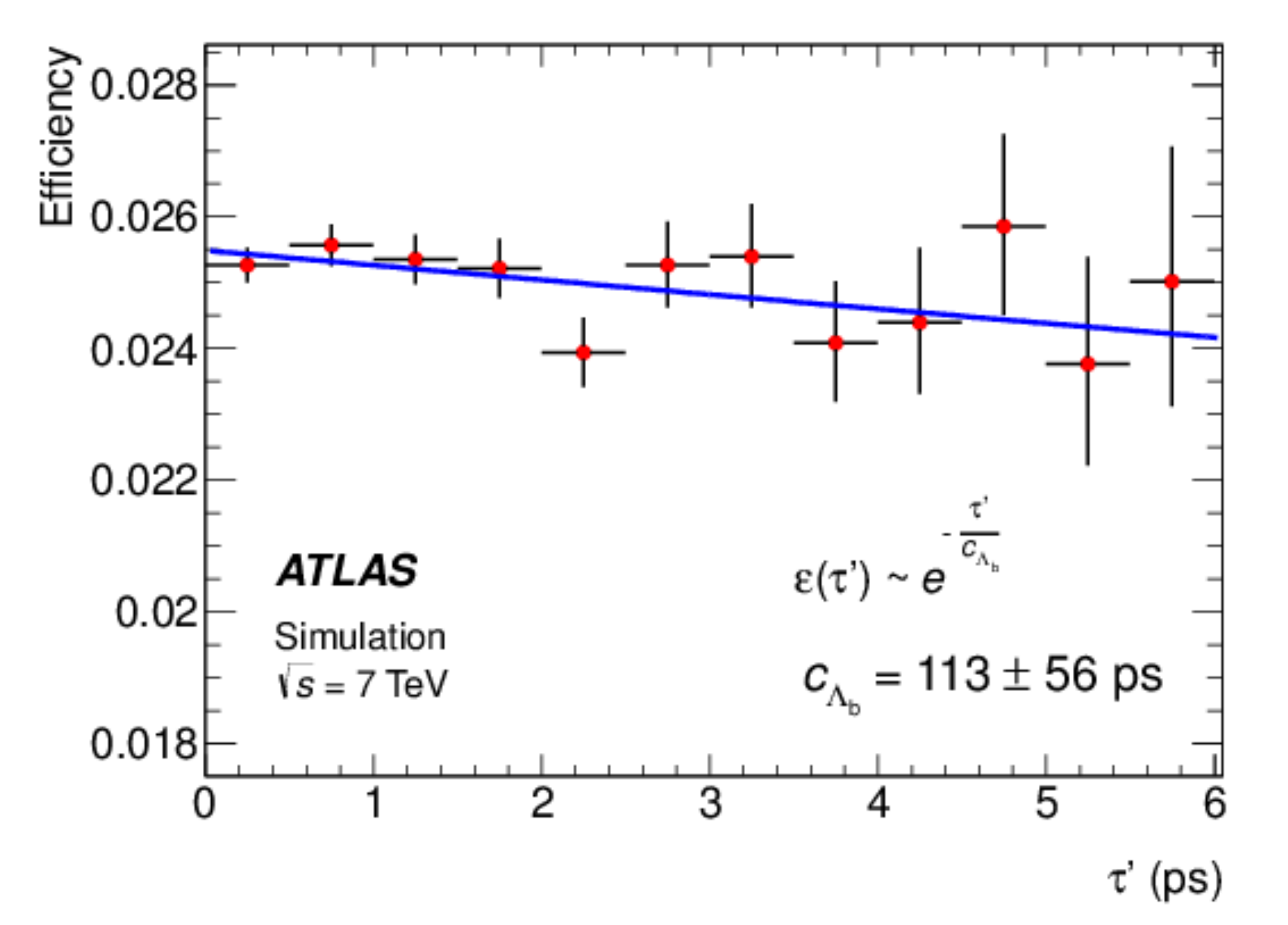}

  \caption{Extraction of the efficiency correction for \LamB. The MC efficiency is fitted with an exponential function. The $y$ axis has a suppressed zero. The displayed error is statistical only.}

  \label{fig:eff_LamB}
\end{center}
\end{figure}

Two sources are responsible for the selection bias in the \LamB\ decay time: the \Vz\ reconstruction efficiency and the trigger selection. The \Vz\ reconstruction efficiency depends on the decay distance from the center of the detector, as tracks from decays further away from the center leave fewer hits in the ID. Since the \LamB\ decay length and the distance of the \Lam\ vertex from the center of the detector are correlated (the latter includes the former), this biases the measured proper decay time toward smaller values. The other source of the bias is the muon trigger, which affects the distribution of the muon transverse impact parameter, $d_0$. Applying the tag-and-probe method to \Jpsi\ decays, the trigger efficiency as a function of $d_0$ is measured for a single-muon trigger in data. The simulation shows that the dimuon-trigger efficiency can be expressed as a product of single-muon efficiencies.

The MC events are reweighted to reproduce the observed trigger bias. The efficiency correction, $\varepsilon(\tau')$, is determined using this weighted MC sample. It is modeled as a simple exponential, $\varepsilon(\tau') \propto e^{-\tau'/c_{\Lambda_b}}$, where $c_{\Lambda_b}$ denotes the slope of the efficiency correction. The exponential form is chosen for $\varepsilon(\tau')$ because it describes the MC well and is particularly easy to convolve with the resolution model. The slope of the exponential, $c_{\Lambda_b}$, is extracted from a fit to the MC decay time efficiency plot shown in \FIG~\ref{fig:eff_LamB}. The extracted value is $c_{\Lambda_b} = 113 \pm 56$~ps; i.e. for a decay time of 6~ps the efficiency decreases by 5\%.

\subsection{Parameters determined from the fit}
\label{section:param}

The full PDF has 12 free parameters:
\BeginParams
\ItemParams the \LamB\ mass and lifetime, $m_{\Lambda_b}$ and $\tau_{\Lambda_b}$;
\ItemParams the fraction of signal events, $f_{\rm sig}$;
\ItemParams the error scale factors, $S_m$ and $S_{\tau}$;
\ItemParams the slope of the mass dependence of the background, $b$;
\ItemParams the pseudolifetimes of the long-lived background, $\tau_{\rm bkg,1}$ and $\tau_{\rm bkg,2}$; the exponential slope of the non-Gaussian prompt background, $\tau_{\rm bkg,3}$; and
\ItemParams the relative fractions of the various background contributions, $f_1$, $f_2$, and $f_3$.
\EndParams

Other quantities are calculated from the fit parameters. The number of signal and background candidates, $N_{\rm sig}$ and $N_{\rm bkg}$, are calculated as $N_{\rm sig} = f_{\rm sig}N$ and $N_{\rm bkg} = (1-f_{\rm sig})N$, where $N$ is the total number of candidates. The mass and proper decay time resolutions are calculated from the fit parameters, too. By analogy with a Gaussian distribution, the mass resolution $\sigma_m$ is defined as half of that mass range for which the integral of $\mathcal{M}_{\rm s}$ retains 68.3\% of the number of signal events symmetrically around the fitted \LamB\ mass. The proper decay time resolution, $\sigma_\tau$, is determined in the same fashion by integrating the prompt background PDF.

% %%%%%%%%%%%%%%%%%%%%%%%%%%%%%%%%%%%%%%%%%%%%%%%%%%%%%%%%%%%%%%%%%
\section{Extraction of the lifetime and mass}
\subsection{Results of the maximum likelihood fit}

\begin{table}[t]
\begin{center}
  \caption{Results of the simultaneous mass and decay time maximum likelihood fit for \LamB. The uncertainties shown are statistical only. The number of degrees of freedom used for the $\chi^2$ calculation is $N_{\rm dof} = 61$.}
  \medskip

  \begin{tabular}{@{}lcccc@{}}
    \hline\hline
    \textbf{Parameter}    & \textbf{Value}        & \quad & \textbf{Par.}         & \textbf{Value}        \\\hline
    $m_{\Lambda_{b}}$     & $5619.7 \pm 0.7$~MeV  &       & $\chi^2/N_{\rm dof}$  & $1.09$                \\
    $\tau_{\Lambda_{b}}$  & $1.449 \pm 0.036$~ps  &       & $N_{\rm sig}$         & $2184 \pm 57$         \\
    $f_{\rm sig}$         & $0.268 \pm 0.007$     &       & $N_{\rm bkg}$         & $5970 \pm 160$        \\
    $S_m$                 & $1.18 \pm 0.03$       &       & $\sigma_m$            & $31.1 \pm 0.8$~MeV    \\
    $S_{\tau}$            & $1.05 \pm 0.02$       &       & $\sigma_\tau$         & $0.117 \pm 0.003$~ps  \\
    \hline\hline
  \end{tabular}

  \label{tab:results_lamB}
\end{center}
\end{table}

\begin{figure}[t]
\begin{center}
  \includegraphics[width=\figWidth\linewidth]{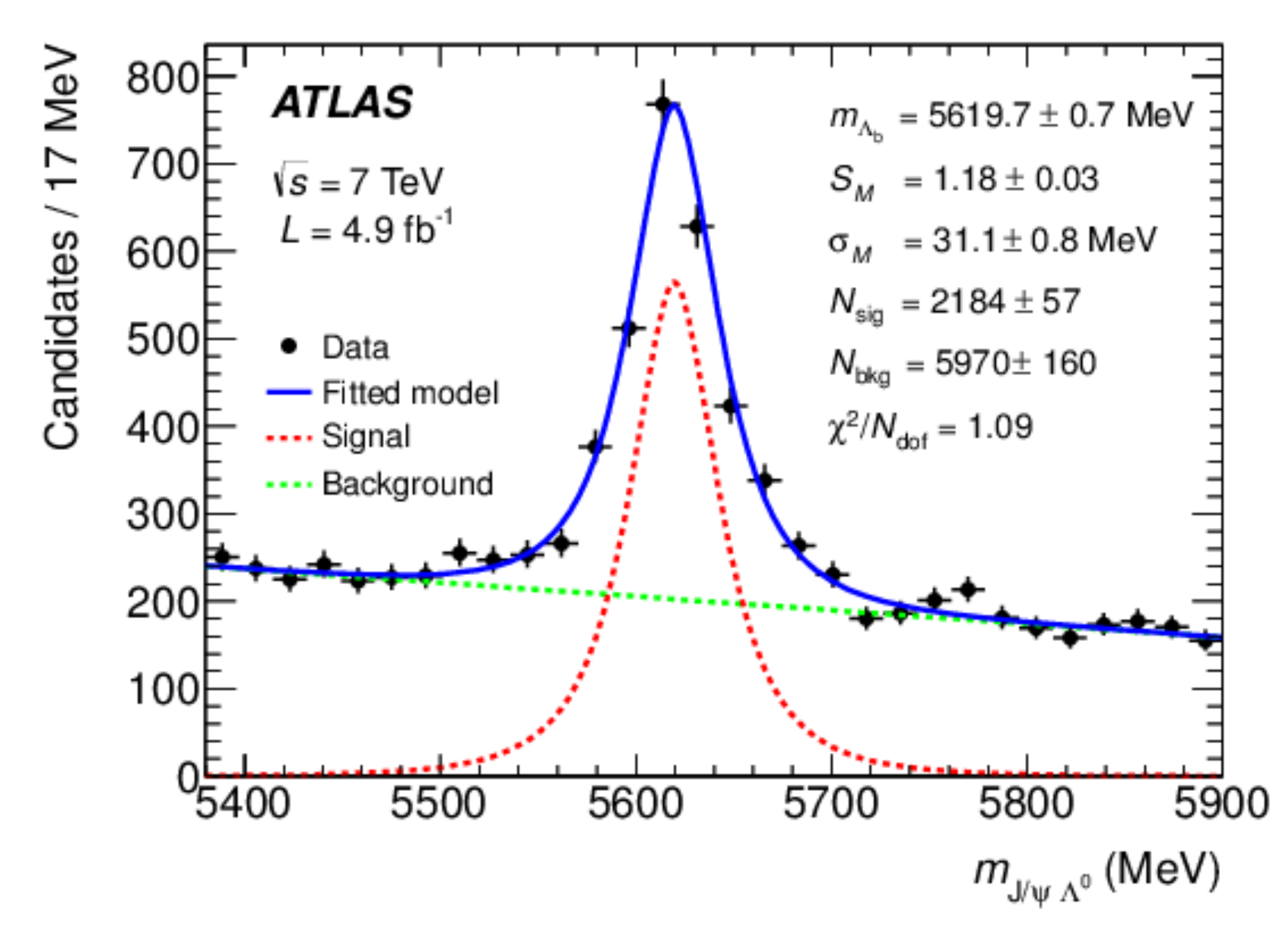}
  \includegraphics[width=\figWidth\linewidth]{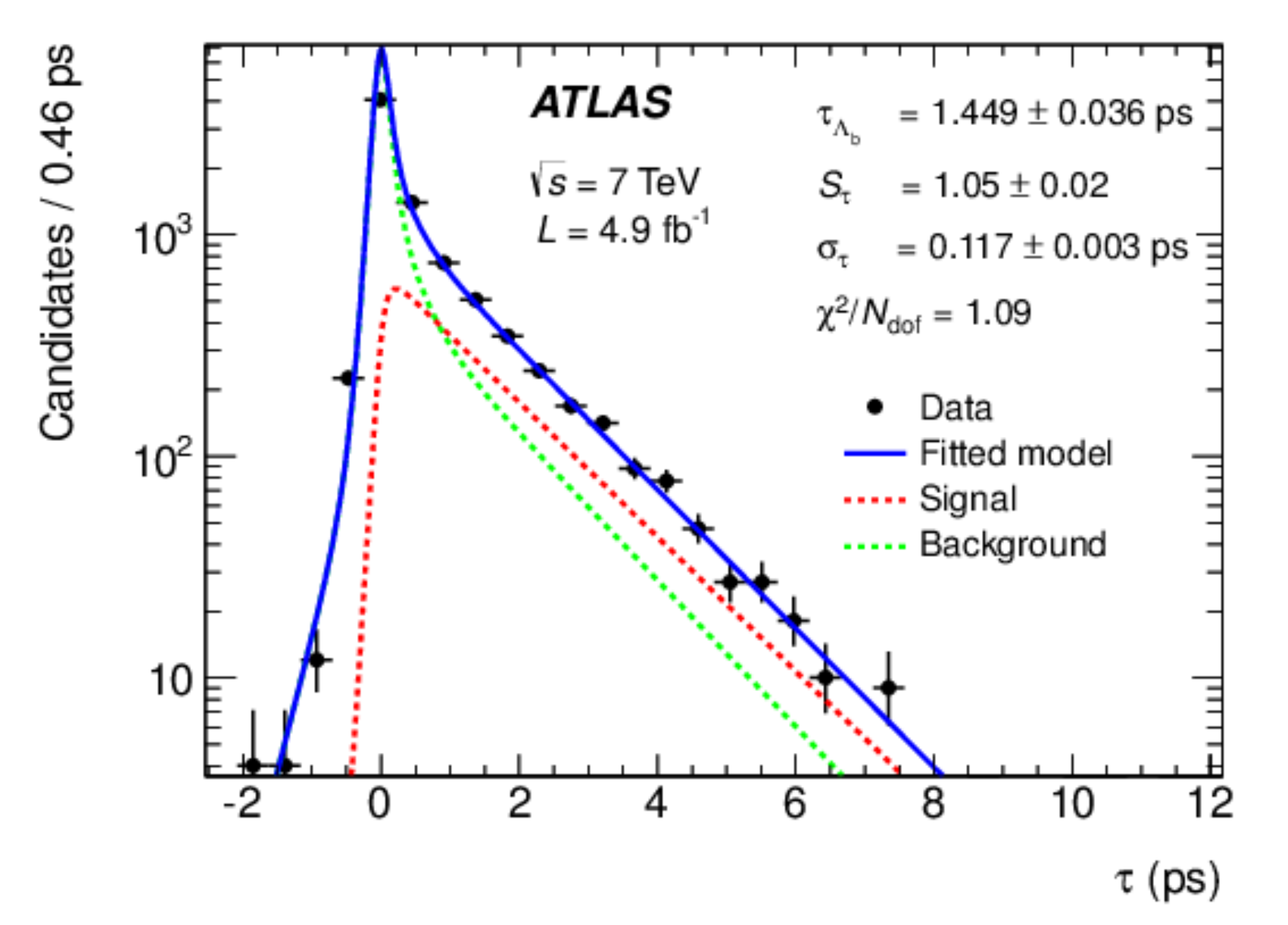}

  \caption{Projections of the fitted PDF onto the mass (\FirstImage) and the proper decay time (\SecondImage) axes for \LamB\ candidates. The errors of the listed fit result values are statistical only. The $\chi^2/N_{\rm dof}$ value is calculated from the data set binned in mass and decay time with the number of degrees of freedom, $N_{\rm dof} = 61$.}

  \label{fig:lifetime_lamB}
\end{center}
\end{figure}

The results of the maximum likelihood fit are listed in \TABLE~\ref{tab:results_lamB}. The table shows only the most important fitted parameters, calculated parameters, and a $\chi^2/N_{\rm dof}$ value which quantifies the fit quality. The $\chi^2/N_{\rm dof}$ value is calculated from the dataset binned in mass and decay time with 61 degrees of freedom. The sizes of the bins are commensurate with the measured mass and decay time resolutions, and only bins with more than 11 entries are used for the $\chi^2$ calculation. This  requirement is imposed so that the error on the number of entries in each bin can be taken as Gaussian. The lifetime result is corrected for the selection bias (see Sec.~\ref{sec:fit}); the size of the correction is $+19$~fs. The estimated correlation between the mass and lifetime is small, 0.002. Projections of the PDF onto the mass and proper decay time axes are shown in \FIG~\ref{fig:lifetime_lamB}.

\subsection{Systematic uncertainties}
\label{sec:syst}

Systematic errors are estimated by changing various parameters of the analysis and observing the shift in the extracted mass and lifetime. The shift with respect to the baseline result is then quoted as a systematic uncertainty. The non-negligible systematic uncertainties are summarized in Table~\ref{tab:syst}. The individual errors are added in quadrature, yielding total systematic errors of the lifetime and mass measurements, $\sigma_\tau^{\rm syst} = 17$~fs and $\sigma_m^{\rm syst} =  1.1$~MeV, respectively. Details of the determination of the systematic uncertainties follow.

\subsubsection{Event selection and reconstruction bias}
  Two effects that lead to a selection bias for \LamB\ candidates as a function of decay time have been identified: the dominant contribution comes from the muon trigger, which slightly biases the transverse impact parameter of muons, $d_0$, toward smaller values. The second bias comes from the \Vz\ reconstruction. The event selection bias is corrected using the MC simulation to determine the efficiency as a function of the decay time as described in Sec.~\ref{sec:fit}. The slope of the efficiency correction function, $c_{\Lambda_b}$, is determined with a statistical accuracy of 50\% (56~ps). Using standard error propagation, the contribution of this uncertainty to the overall error is evaluated to be 9~fs.

  Since the two bias corrections rely on MC simulation, systematic uncertainties due to the corrections are estimated. These are determined separately for the \Vz\ reconstruction bias and for the trigger bias. In the MC simulation the bias correction shifts the \LamB\ lifetime by 26~fs, of which 10~fs are due to the \Vz\ reconstruction and 16~fs are due to the trigger requirement. The systematic error due to the \Vz\ bias correction is estimated by varying the \LamB\ transverse momentum in the MC simulation (using kinematic reweighting) to probe the \pt\ dependence of the correction. The magnitude of the variation is about 3 times the difference between the mean \pt\ in data and the MC simulation. The corresponding error is estimated to be 4~fs. To ensure a good description of the trigger bias, the MC sample is reweighted using single-muon-trigger efficiencies expressed as a function of muon $d_0$ values that are extracted from data. The weighting functions are parameterized as linear functions, $w(d_0) \propto 1+a d_0$, and their slope $a$ is determined by a linear fit in bins of the measured muon \pt\ and \eta. To assess the systematic error on the trigger bias correction, the weighting parameters $a$ are varied by their errors. This produces a lifetime shift of 7~fs, which is used as a systematic error. The total systematic error calculated as a quadratic sum of the individual contributions is 12~fs.

  To assess the systematic error in the mass measurement due to the event selection, the MC distribution of $\Delta m = m^{\rm MC} - m$, where $m^{\rm MC}$ is the generated mass, is fitted with a double Gaussian. The systematic error, given by the shift of the mean of the double Gaussian, is estimated to be 0.9~MeV. The mass shift is caused by the muon-trigger \pt\ thresholds: muons with larger \pt\ have higher probability of being selected than low-\pt\ muons. As a consequence, muons whose \pt\ is mismeasured as larger than the true value have a higher probability of being reconstructed than muons whose \pt\ is mismeasured as smaller, which creates a small asymmetry of the mass peak.

\subsubsection{Background fit models}
  Alternative background models are used to assess the sensitivity of the results to the choice of background parametrization. A second-order polynomial and an exponential mass dependence of the $\cal{M}_{\rm b}$ PDF are tested. In addition the decay time dependence is modified by adding a third exponential into the nonprompt background component, $\cal{T}_{\rm np}$. The alternative background PDFs fit the data well. These changes result in a lifetime shift of 2~fs and a mass shift of 0.2~MeV. In the fit model the decay time and mass are assumed to be uncorrelated. To test this assumption the fit's mass range limits, $m_{\rm min}$ and $m_{\rm max}$, are varied independently by 60~MeV. This changes the relative contribution of the background from  the left and right sidebands, and the mass and lifetime are extracted again for these new mass ranges. While the change of $m_{\rm max}$ has a minimal impact on the extracted mass and lifetime, the change of $m_{\rm min}$ produces a lifetime shift of 9~fs. This value is added to the total systematic error due to background modeling.

\subsubsection{\Bd contamination}
  The number of \Bd\ candidates misidentified as \LamB\ is estimated by a fit to the mass distribution of the candidates which fall in the \LamB\ signal region, $5.52~{\rm GeV} < m_{J/\psi \Lambda^0} < 5.72$~GeV, under the hypothesis that they are \BdDecay\ decays. A fit to a Gaussian peak on a linear background yields $82 \pm 46$ \Bd\ candidates. Since these candidates are treated as \LamB, their pseudolifetime is scaled up by the ratio of the \LamB\ and \Bd\ masses, $\tau_{B_d}^* = \tau_{B_d} m^{\rm PDG}_{\Lambda_b}/m^{\rm PDG}_{B_d} = 1617$~fs (the decay time change due to the difference in \pt\ reconstructed under the two hypotheses is negligible). If all such background candidates contribute to the fitted \LamB\ lifetime, it would cause a shift of 7~fs. This is quoted as a conservative estimate of the systematic error. The error on the mass measurement is estimated by relaxing the $\mathcal{P}_{\Lambda^0_b} - \mathcal{P}_{B_d}$ cut to double the estimated \Bd\ background. This results in a \LamB\  mass shift of 0.2~MeV.

\subsubsection{Residual misalignment of the ID}
  The distribution of the transverse impact parameter, $d_0$, of tracks originating from the PV is used to estimate the geometrical distortions due to residual misalignment. The geometry in the MC simulation is distorted by adjusting the positions of the ID modules so that the $d_0$ of tracks coming from the PV is biased by the same amount as observed in data. The mass-lifetime fit is performed with simulated data using the default (ideal) geometry and the sample with geometry distortions. A shift of 1~fs is observed between the two measurements and is assigned as a systematic error due to residual misalignment. No significant mass shift is observed.

\subsubsection{Uncertainty in the amount of ID material}
  Inaccurate modeling of the amount of material in the ID could affect the measurement since the tracking algorithm estimates the particle energy loss using a material map. To explore this uncertainty, the MC simulation is repeated with 20\% more material in the ID silicon detectors (Pixel and SCT) and their supporting services, which is large compared to the estimated uncertainty of 6\%(9\%) in the Pixel (SCT) detectors (see Ref.~\cite{Aad:2011cxa}). The resulting shifts of 3~fs in lifetime and 0.2~MeV in mass are conservative estimates of the systematic uncertainties from this source.

\subsubsection{Uncertainty in the tracking momentum scale}
  The \Ks\ mass value is used to estimate the uncertainty in the track momentum determination. The \Ks\ mass extracted from a fit to the invariant mass agrees with the PDG's world average within 0.03\%. Such a shift corresponds to a track momentum scale shift of 0.05\%. The momentum scale can be further tested using the reconstructed \Jpsi\ mass. The observed mass shift corresponds to a momentum scale error of $-0.03$\%, in agreement with the assumption of $\pm$0.05\%. Shifting the momenta of all tracks in the MC simulation by this amount yields a \LamB\ mass shift of 0.5~MeV. No significant lifetime shift is observed.

\subsubsection{Other systematic errors}
  Other sources of systematic errors are investigated, such as an alternative choice of the PV (e.g. using the collision vertex whose tracks have the highest sum of $p_{\rm T}^2$) and the use of a PDF with the error distributions modeled separately for signal and background. These changes do not result in a significant mass or lifetime shift.

\begin{table}[t]
\begin{center}
  \caption{Summary of the systematic uncertainties of the lifetime measurement, $\sigma_\tau^{\rm syst}$, and the mass measurement, $\sigma_m^{\rm syst}$, for \LamB. }
  \medskip
  \begin{tabular}{@{}lcc@{}}
    \hline\hline
    \textbf{Systematic uncertainty} & \boldmath{$\sigma_\tau^{\rm syst}$ \textbf{(fs)}} & \boldmath{$\sigma_m^{\rm syst}$} \textbf{(MeV)} \\\hline
      Selection/reco. bias          &  12        & 0.9  \\
      Background fit models         &   9        & 0.2  \\
      \Bd\ contamination            &   7        & 0.2  \\
      Residual misalignment         &   1        &  $\cdots$   \\
      Extra material                &   3        & 0.2  \\
      Tracking \pt\ scale           &   $\cdots$ & 0.5  \\
    \hline
      Total systematic error        &  17        & 1.1  \\
    \hline\hline
  \end{tabular}

  \label{tab:syst}
\end{center}
\end{table}

\subsection{Cross-check with \boldmath{\BdDecay}}
\label{sec:bd_results}

The \BdDecay\ channel has the same decay topology as \LamBDecay\ and can be used to cross-check the \LamB\ results and to determine the ratio of the \LamB\ and \Bd\ lifetimes. The analysis and systematic error studies, described in the previous sections, are repeated for the \Bd. The \Bd\ channel is subjected to exactly the same kinematic cuts as for the \LamB\ channel and therefore has similar systematic effects. The mass range used for the \Ks\ pre-selection is $440\ {\rm MeV} < m_{\pi^+\pi^-} < 570\ {\rm MeV}$, and the \Bd\ invariant mass must lie in the range $5.1\ {\rm GeV} < m_{J/\psi K_S^0} < 5.5\ {\rm GeV}$. This selection is chosen to be as close as possible to the \LamB\ selection rather than to maximize the \Bd\ signal yield. Using the maximum likelihood fit, the \Bd\ lifetime and mass are measured to be $\tau_{B_d} = 1.509 \pm 0.012({\rm stat}) \pm 0.018({\rm syst})\ {\rm ps}$ and $m_{B_d} = 5279.6 \pm 0.2({\rm stat}) \pm 1.0({\rm syst})\ {\rm MeV}$. These values are consistent with the world averages, $\tau_{B_d}^{\rm PDG} = 1.519 \pm 0.007$~ps and $m_{B_d}^{\rm PDG} = 5279.50 \pm 0.30$~MeV \cite{PDG}.

\section{Results and conclusions}

The \LamB\ lifetime and mass are measured to be
\begin{gather}
  \tau_{\Lambda_b} = 1.449 \pm 0.036({\rm stat}) \pm 0.017({\rm syst})\ {\rm ps}, \notag\\
  m_{\Lambda_b}    = 5619.7 \pm 0.7({\rm stat}) \pm 1.1({\rm syst})\ {\rm MeV}. \notag
\end{gather}
These results agree with the world average values of the \LamB\ lifetime, $\tau_{\Lambda_{b}}^{\rm PDG} = 1.425 \pm 0.032$~ps and mass, $m_{\Lambda_{b}}^{\rm PDG} = 5619.4 \pm 0.7$~MeV \cite{PDG}, and with a recent determination of the \LamB\ mass by the LHCb experiment, $m_{\Lambda_{b}}^{\rm LHCb} = 5619.19 \pm 0.70({\rm stat}) \pm 0.30({\rm syst})$~MeV \cite{Aaij2012241}.
The ratio of the \LamB\ and \Bd\ lifetimes is
\begin{gather}
  R = \tau_{\Lambda_b}/\tau_{B_d} = 0.960 \pm 0.025({\rm stat}) \pm 0.016({\rm syst}).\notag
\end{gather}
The statistical and systematic errors are propagated from the errors of the lifetime measurements. The systematic errors are conservatively assumed to be uncorrelated. This value is between the recent determination by D\O, $R^{\rm D\O} = 0.864 \pm 0.052({\rm stat}) \pm 0.033({\rm syst})$ \cite{Abazov:2012iy}, and the measurement by CDF, $R^{\rm CDF} = 1.020 \pm 0.030({\rm stat}) \pm 0.008({\rm syst})$ \cite{Aaltonen:2010pj}. It agrees with the heavy quark expansion calculations which predict the value of the ratio between $0.88$ and $0.97$~\cite{HQET} and is compatible with the next-to-leading-order QCD predictions with central values ranging between $0.86$ and $0.88$ (uncertainty of $\pm0.05$) \cite{NLO}.

%
% %%%%%%%%%%%%%%%%%%%%%%%%%%%%%%%%%%%%%%%%%%%%%%%%%%%%%%%%%%%%%%%%%

% Acknowledgements for papers with collision data
% Version 9-Feb-2012

\section*{Acknowledgements}

We thank CERN for the very successful operation of the LHC, as well as the
support staff from our institutions without whom ATLAS could not be
operated efficiently.

We acknowledge the support of ANPCyT, Argentina; YerPhI, Armenia; ARC,
Australia; BMWF, Austria; ANAS, Azerbaijan; SSTC, Belarus; CNPq and FAPESP,
Brazil; NSERC, NRC and CFI, Canada; CERN; CONICYT, Chile; CAS, MOST and NSFC,
China; COLCIENCIAS, Colombia; MSMT CR, MPO CR and VSC CR, Czech Republic;
DNRF, DNSRC and Lundbeck Foundation, Denmark; EPLANET and ERC, European Union;
IN2P3-CNRS, CEA-DSM/IRFU, France; GNAS, Georgia; BMBF, DFG, HGF, MPG and AvH
Foundation, Germany; GSRT, Greece; ISF, MINERVA, GIF, DIP and Benoziyo Center,
Israel; INFN, Italy; MEXT and JSPS, Japan; CNRST, Morocco; FOM and NWO,
Netherlands; RCN, Norway; MNiSW, Poland; GRICES and FCT, Portugal; MERYS
(MECTS), Romania; MES of Russia and ROSATOM, Russian Federation; JINR; MSTD,
Serbia; MSSR, Slovakia; ARRS and MVZT, Slovenia; DST/NRF, South Africa;
MICINN, Spain; SRC and Wallenberg Foundation, Sweden; SER, SNSF and Cantons of
Bern and Geneva, Switzerland; NSC, Taiwan; TAEK, Turkey; STFC, the Royal
Society and Leverhulme Trust, United Kingdom; DOE and NSF, United States of
America.

The crucial computing support from all WLCG partners is acknowledged
gratefully, in particular from CERN and the ATLAS Tier-1 facilities at
TRIUMF (Canada), NDGF (Denmark, Norway, Sweden), CC-IN2P3 (France),
KIT/GridKA (Germany), INFN-CNAF (Italy), NL-T1 (Netherlands), PIC (Spain),
ASGC (Taiwan), RAL (UK) and BNL (USA) and in the Tier-2 facilities
worldwide.

% %%%%%%%%%%%%%%%%%%%%%%%%%%%%%%%%%%%%%%%%%%%%%%%%%%%%%%%%%%%%%%%%%

%

\onecolumngrid
\clearpage

% ATLAS Collaboration author list for 18-MAY-2012
% Data extracted on 06-Jun-2012 for paperid 232
% \documentclass[11pt]{article}
% \usepackage{a4wide}\begin{document}
\begin{flushleft}
{\Large The ATLAS Collaboration}

\bigskip

G.~Aad$^{\rm 48}$,
B.~Abbott$^{\rm 111}$,
J.~Abdallah$^{\rm 11}$,
S.~Abdel~Khalek$^{\rm 115}$,
A.A.~Abdelalim$^{\rm 49}$,
O.~Abdinov$^{\rm 10}$,
R.~Aben$^{\rm 105}$,
B.~Abi$^{\rm 112}$,
M.~Abolins$^{\rm 88}$,
O.S.~AbouZeid$^{\rm 158}$
% ...
\bigskip

% ...
$^{10}$ Institute of Physics, Azerbaijan Academy of Sciences, Baku, Azerbaijan\\
$^{11}$ Institut de F\'isica d'Altes Energies and Departament de F\'isica de la Universitat Aut\`onoma  de Barcelona and ICREA, Barcelona, Spain\\
% ...
$^{48}$ Fakult\"{a}t f\"{u}r Mathematik und Physik, Albert-Ludwigs-Universit\"{a}t, Freiburg i.Br., Germany\\
$^{49}$ Section de Physique, Universit\'e de Gen\`eve, Geneva, Switzerland\\
% ...
$^{88}$ Department of Physics and Astronomy, Michigan State University, East Lansing MI, United States of America\\
% ...
$^{105}$ Nikhef National Institute for Subatomic Physics and University of Amsterdam, Amsterdam, Netherlands\\
% ...
$^{111}$ Homer L. Dodge Department of Physics and Astronomy, University of Oklahoma, Norman OK, United States of America\\
$^{112}$ Department of Physics, Oklahoma State University, Stillwater OK, United States of America\\
% ...
$^{115}$ LAL, Universit\'e Paris-Sud and CNRS/IN2P3, Orsay, France\\
% ...
$^{158}$ Department of Physics, University of Toronto, Toronto ON, Canada\\
% ...
\end{flushleft}

\end{document}